# Visualizing superconductivity in an inversion-symmetry- broken doped Weyl semimetal

Zhenyu Wang[1,2], Jorge Olivares[2], Hiromasa Namiki[3], Vivek Pareek[4], Keshav Dani[4], Takao Sasagawa[3], Vidya Madhavan[2*] and Yoshinori Okada[4*]

[1]Department of Physics and Key Laboratory of Strongly-Couple Quantum Matter Physics, CAS, University of Science and Technology of China, Hefei, Anhui 230026, China

[2]Department of Physics and Frederick Seitz Materials Research Laboratory, University of Illinois Urbana-Champaign, Urbana, Illinois 61801, USA

[3]Materials and Structures Laboratory, Tokyo Institute of Technology, Kanagawa 226-8503, Japan

[4]Okinawa institute of Science and Technology, Okinawa Prefecture 904-0412, Japan

**The Weyl semimetal MoTe$_2$ offers a rare opportunity to study the interplay between Weyl physics and superconductivity. Recent studies have found that Se substitution can boost the superconductivity up to 1.5K, but suppress the T$_d$ structure phase that is essential for the emergence of Weyl state. A microscopic understanding of possible coexistence of enhanced superconductivity and the T$_d$ phase has not been established so far. Here, we use scanning tunneling microscopy (STM) to study a optimally doped new superconductor MoTe$_{1.85}$Se$_{0.15}$ with bulk Tc ~ 1.5K. By means of quasiparticle interference imaging, we identify the existence of low temperature T$_d$ phase with broken inversion symmetry where superconductivity globally coexists. Consistently, we find that the superconducting coherence length, extracted from both the upper critical field and the decay of density of states near a vortex, is much larger than the characteristic length scale of existing dopant derived chemical disorder. Our findings of robust superconductivity arising from a Weyl semimetal normal phase in MoTe$_{1.85}$Se$_{0.15}$, makes it a promising candidate for realizing topological superconductivity.**

## Introduction

A Weyl semimetal (WSM) is a topological, gapless system which hosts three-dimensional Weyl cones in the bulk and Fermi arc states on the surface[1-6]. It occurs in a system breaking either time-reversal or inversion symmetry, where nondegenerate conduction and valence bands intersect at arbitrary points in momentum space, forming pairs of the so-called 'Weyl nodes' with opposite spin chirality. Near these nodes, the low-energy excitations can be described as linearly dispersing Weyl-Fermions. When a WSM falls into superconducting state, it may lead to a branch of exotic topological phenomena such as Weyl superconductivity[7,8], Fulde-Ferrell-Larkin-Ovchinnikov (FFLO) paring states[9-11] and time-reversal invariant topological superconductivity[12-14]. This has attracted much attention recently as it may provide a promising route to realize zero energy modes related to Majorana Fermions[15], which may be useful for topological quantum computation. The experimental search for topological superconductor phases have been recently accelerated further by experimental discovery of edge supercurrent in $MoTe_2$[16] and higher-order hinge state in Josephson junction interference from isostructural $WTe_2$[17].

In this work we present microscopic studies of a doped superconducting Weyl semimetal, $Mo(Te, Se)_2$. The parent compound, $MoTe_2$, undergoes a structural transition from a 1T' phase to a low temperature $T_d$ phase[18] at Ts~250 K. The difference between these two phases is a change of the stacking angle from monoclinic (93.9°) to orthorhombic (90°), as shown in Fig. 1a. The $T_d$ phase breaks inversion symmetry which is a necessary condition for the emergence of a type II WSM[19,20]. Experimentally, the existence of topological Fermi arc states has been supported by a series of photoemission spectroscopy measurements[21-24]. $T_d$-$MoTe_2$ becomes superconducting below 0.1K at ambient pressure, and an edge supercurrent has recently been revealed in the oscillation of the critical current versus magnetic field[16]. Interestingly, external pressure and chemical substitution can significantly enhance the superconducting transition temperature ($T_c$) from 0.1K to several Kelvin[25-30]. However, previous studies for both cases show the phase diagram representing rapid $T_c$ enhancement across doping level where the structural transition from T' to $T_d$ almost disappears[29-31]. Thus, microscopic understanding of how the enhanced superconductivity may coexist with suppressed $T_d$ phases becomes an essential question. While chemical substitution case $MoTe_{2-x}Se_x$ provides a platform for detailed investigation by spectroscopic scanning tunneling microscopy (STM), microscopic picture of emergent superconductivity in $MoTe_{2-x}Se_x$ has not been established so far.

The schematic phase diagram for $MoTe_{2-x}Se_x$ is shown in figure 1b [32]. With increasing Se, $T_c$ increases while the $T_d$ phase is known to be suppressed at larger Se concentrations. Here, our main focus is the superconducting compound with x=0.15 since it could be the optimal composition to search for the coexistence of enhanced superconductivity and $T_d$ phase. We use low-temperature STM to pinpoint the low temperature structural phase of substituted $MoTe_{1.85}Se_{0.15}$ by means of quasiparticle interference imaging and provide the first evidence that the enhanced superconductivity and the $T_d$ structural phase can coexist at the microscopic scale.

## Methods

Single crystals of $Mo(Te, Se)_2$ were grown by the chemical vapor transport method. The samples were characterized by XRD and XRF to confirm systematic doping without phase segregation. The temperature dependence of resistivity was measured by PPMS. The samples were cleaved *in-situ* at 90K, and then immediately transferred into STM head (custom Unisoku 1300). Chemically etched tungsten tips were used in all the measurements, after being checked on a clean Cu (111) surface. All dI/dV measurements were taken using a standard lock-in technique with ~0.07meV peak to peak modulation, at a frequency of 987.5Hz.

**Results**

$MoTe_{1.85}Se_{0.15}$ samples are first characterized by the resistivity measurements. The 1T' to $T_d$ bulk structural transition is revealed by a resistivity anomaly which is seen as a hysteresis in the resistivity upon cooling or warming around 250 K as can be seen for the parent compound (Fig. 1c). In $MoTe_{1.85}Se_{0.15}$, this anomaly around 250 K becomes relatively weak but still visible. Importantly, Se substitutions to the Te sites also effectively enhance superconductivity, as the onset of the $T_c$ occurs around 3 K and zero resistivity can be achieved around 1.5 K (inset of Fig. 1c).

$Mo(Se,Te)_2$ consists of van der Waals stacking of (Se-Te)-Mo-(Se-Te) sandwich layers. Cleaving naturally occurs between two of the stacking layers, resulting in a Te-terminated surface. Figure 1d shows a typical atomic-resolution topographic image of the cleaved $MoTe_{1.85}Se_{0.15}$. Two inequivalent atomic rows are visible, with one slightly higher than the other. Depressions can be observed in both atomic rows, which corresponds to the substituted Se atoms that have a smaller radius. To study the effect of Se doping on the normal state electronic structure, we characterize the local density of states by differential tunneling conductance (dI/dV (r, eV)) measurements. Figure 1e shows tunneling spectra averaged over 50nm*50nm regions on pristine and on Se substituted samples. They have similar overall spectral shapes, both exhibiting semi-metallic behavior. The spectrum acquired on the pristine sample shows a pronounced peak at E ~ 550 meV and a hump feature at 320 meV, which we associate with the top of two bulk bands as in the prior band structure calculations[33,34]. By comparison, these two features move to 430 and 220 meV respectively in the Se substituted sample, indicating a shift of the chemical potential to higher energies. Thus, we conclude that the $MoTe_{1.85}Se_{0.15}$ samples are effectively electron doped compared to the pristine samples. We note that since Se is naively expected to be isovalent, further work needs to be done to identify the origin of the electron doping. As highlighted by dashed circle in Figure 1d, we observe topographic corrugation, which is presumably related to the effect of dopant Se.

To check the possible phase coexistence between 1T' and $T_d$ at low temperature, we have performed Raman spectroscopy using exfoliated flake of $MoTe_{1.85}Se_{0.15}$ (Supplemental Material note 1, SM 1 [35]). The thickness of exfoliated flake is ~20 nm, which is thick enough to provide bulk information. As shown in Fig. 1e, Raman spectra at 295 and 77K show large amount of spectral weight between 120 $cm^{-1}$ and 135 $cm^{-1}$. Within this energy window, based on previous report on pristine $MoTe_2$ [36], two characteristic Raman shifts have been assigned to the $T_d$ phase (127.5 $cm^{-1}$ and 130.8 $cm^{-1}$), in addition to one Raman shift for T' phase (128.3 $cm^{-1}$). These characteristic Raman shifts are indicated by vertical lines in Fig. 1f-g. For < 130 $cm^{-1}$, two Raman shifts from $T_d$ and T' phases exist very close with each other. Because of this reason in this energy region, separating peaks becomes difficult. On the other hand, we observe $T_d$ phase originated clear peak around 132 $cm^{-1}$ at 77K (indicated by arrow in Fig. 1f). Consistent with phase diagram (Fig. 1b), enhanced stability of $T_d$ phase with cooling sample is more clearly visualized by the normalized Raman spectra I(77K)/I(295K) around 132 $cm^{-1}$. This further confirms peak around 132 $cm^{-1}$ is not from extrinsic background. Thus, our bulk structure characterization suggests the existence of robust $T_d$ phase coexisting with T' phase. This is presumably related to corrugation seen in topographic image (dashed line in Fig. 1d). Hereafter, our main goal is to search for possible enhanced superconductivity compared to that in pristine sample, which arises from the $T_d$ component.

At 0.3K, we observe clear coherence peaks form at ±0.2meV in the tunneling spectra with a suppression of spectral weight near Fermi energy, which we identify to be the superconducting gap. Despite the presence of Se dopants which make our system chemically disordered, as shown in Fig. 2a and b, the gap feature is quite homogeneous over a large region, and we find similar spectra on all the surfaces that have been measured. This homogeneously extended superconducting phase is an essential characteristic for the potential realization of global topologically non-trivial quantum states.

To further characterize superconductivity and its coherence length, both temperature and magnetic field dependence of superconducting gap are measured. The temperature dependence of the tunneling spectra is shown in Fig. 2c. One can see that the superconducting gap is gradually suppressed at elevated temperatures and eventually disappears above 0.75 K. To quantitatively evaluate the gap value, we fit the data with an isotropic s-wave gap and show the temperature dependence of superconducting gap in fig. 2d which exhibits BCS-like behavior (SM 2 [35]). The gap value is extracted to be 0.18 meV, which yields a ratio of $2\Delta/k_BT_c$ ~ 4.5. This value exceeds that for a weak coupling BCS superconductor but is consistent with the one reported in the pressure induced superconductivity of $MoTe_2$[28]. We also carried out spectroscopy studies under a magnetic field perpendicular to the surface. With increasing fields, the superconducting gap feature fades and disappears above 0.15 T (fig. 2e), which predicts $\Delta(H) = \Delta(0)[1-(H/H_{c2})^2]^{1/2}$, resulting in an upper critical field about 0.13± 0.02 T (fig. 2f). Based on Ginzburg-Landau expression $H_{c2} = \Phi_0/2\pi\xi_{ab}^2$, we calculate the superconducting coherence length as $\xi = 51 \pm 5 nm$. If $\xi$ were comparable to the length scale of chemical disorder, as in cuprates[37], one might expect an inhomogeneous gap size in real space. However, consistent with our observations of a homogeneous gap, $\xi$ in our system is much longer than the length scale of chemical disorder.

The superconducting coherence length is further confirmed by imaging vortex cores. The vortices are directly imaged by mapping the conductance inside the gap with an applied magnetic field of 0.04 T. Based on the single magnetic flux quanta $2.07 \times 10^{-15}$ Wb, one should see about 2 vortices in a 350nm X 350nm field of view. Figure 3a shows such a map at zero energy, in which one vortex emerges at the top-left corner with an isotropic shape while other two are partly included. Inside the vortex core, we find a flat DOS without any clear peak-like features (Fig.3b and c). We will discuss this featureless vortex core state later. To extract the superconducting coherence length, we first calculate the averaged zero bias conductance as a function of distance from the core, and then fit the result with an exponential decay $g(r) = g_0 + A exp(-r/\xi)$, where $g_0$ is the normalized zero energy conductance far away from the vortex core. As shown in Fig. 3d, we find $\xi = 62 \pm 5nm$. This value is consistent with the value estimated from $H_{C2}$ ($\xi = 51 \pm 5nm$). The main question is whether the enhanced superconductivity coexists with the $T_d$ phase in the doped samples, which is particularly important as the Weyl physics require the breaking of inversion symmetry. Note that estimated Tc of ~0.75 K on the surface [Fig. 2(d)] is much lower than expected from bulk value 1.5 K (Fig. 1c inset). Largely modulated surface superconductivity from bulk suggests interesting phenomena on surface.

We next present that observation of two types of QPI patterns supports existence of $T_d$ phase, coexisting with superconductivity at low temperature in our doped surface. Relying on the existence of an `arc-like' signature in the quasiparticle interference is perilous since calculations as well as experiments have shown that a trivial surface state may also have the same visual signature[34]. However, it is known that observation of two types of QPI pattern is one of the promising spectroscopic ways to know existence of Td phase. Previous work has shown that the broken inversion symmetry leads to two inequivalent surfaces (Fig. 4a and SM 3 [35]), which can be detected by QPI measurements[34,38]. We have measured QPI patterns from more than five areas, which are separated by distance longer than 10 μm. We emphasize that superconducting gap is homogeneously distributed within a single area to get QPI pattern. Interestingly, we find QPI patterns can be categorized into two groups (SM 3 [35]). Several representative Fourier transforms are shown in Fig. 4e and f, where the corresponding dI/dV maps are measured in field of views up to 100 nm by 100 nm. We report two important observations here. First, based on constant energy contour reported for MoTe$_2$ (Fig. 4d), we can understand the scattering channels as follows. On surface A, we can identify multiple scattering channels: $Q_1$ represents the intra-scattering of the trivial surface states, $Q_2$ is induced by the scattering between surface states and the electron pockets the bulk electron pockets, while $Q_3$ connects the surface states and the projected bulk states at the $\bar{Y}$ points, similar to that in the pristine samples. Second, the strength of these scattering processes differs on the two surfaces. The characteristic difference in QPI patterns between two surfaces are particularly clear at positive energies (highlighted by dashed circles in Fig.4e and f). While it is known that in STM experiments the change of tip condition may alter taste of QPI pattern, the systematic differences in the observed QPI are robust enough to conclude the existence of qualitatively two different types of surfaces in our sample.

Indeed, our observation of two qualitatively different types of QPI patterns are quite analogous to what has been seen in pristine MoTe$_2$ samples, which is well recognized to host $T_d$ phase at low temperature. To show this analogy, we make a direct comparison of the QPI data of pristine MoTe$_2$ and MoTe$_{1.85}$Se$_{0.15}$ in Fig. 5. Here, the data of pristine MoTe$_2$ is from our previous work[34]. By simply shifting the chemical potential of about 50 meV to higher energy, we find clear analogy for both QPI patterns of surfaces A and B in MoTe$_{1.85}$Se$_{0.15}$ to those of pristine MoTe$_2$. Note that this energy shift is qualitatively consistent with that extracted from the dI/dV spectra (Fig. 1e). Thus, we further convince existence of $T_d$ phase on the surface of doped sample.

In principle, if T' and Td phases coexist, totally three types of QPI patterns are possible. They include one pattern from T' phase and the other two patterns from $T_d$ phase with different crystalline polarity. Within a limitation of experimental resolution, our measurements were difficult to detect three types of QPI patterns. While it is possible to see step edges across which different QPI pattern shows up due to different polarity in $T_d$ phase, we have not found such locations in our experiments. However, we emphasize that T' phase alone fails to explain our experimental observation of two types of QPI. It is important to emphasize again that observed length scale of weak topographic corrugation (dashed enclosed line in Fig. 1d), which is presumably related to Se doing and its impact for structural distortion, is less than the spatial extension of Cooper pair wave function and size of area for QPI measurements. Thus, by making an analogy with the different electronic structures observed on the two inequivalent surfaces of MoTe$_2$, we conclude that superconductivity coexists with the $T_d$ phase at the microscopic level in our doped samples, regardless of details of coexistence with T' phases.

**Discussion**

In the presence of broken inversion symmetry, superconductivity could be topologically non-trivial. If the superconducting state also breaks time reversal symmetry, Mo(Te,Se)$_2$ may become a Weyl superconductor which hosts accidental nodes in the gap function[39]. With time reversal symmetry preserved, Weyl superconductivity cannot be realized. Theoretical calculations, however, suggest that inversion symmetry breaking order could lead to two superconducting instabilities, one is in BCS s-wave paring channel and the other one in the odd-parity pairing channel[13,14,39]. The latter will be favored when electrons are strongly correlated[13,39], resulting in a novel topological superconductivity. Recently, there is growing evidence from photoemission measurements showing that the on-site Coulomb interaction plays an important role in $T_d$-MoTe$_2$ and leads to a Lifshitz transition in the band structure[40, 41].

Theoretically, it has been shown that the topologically non-trivial superconducting order parameter must have a sign change between different Fermi surfaces in a time-reversal-invariant Weyl semimetal[12]. Recent muon spin rotation and magneto transport measurements provide clues of s± paring in high pressure and S substituted $MoTe_2$, respectively[27, 42]. In STM measurements, the sign-change order parameter can be probed via the scattering of cooper pairs by non-magnetic defects, which creates quasiparticle bound states inside the gap. We measured the tunneling spectra near surface adatoms (likely to be Cu atom from the tip) and Te vacancies, but no noticeable differences were detected within the gap (SM 4 [35]). Usually, the observation of in-gap states in a s± pairing superconductor requests a moderate scattering potential [43]. If the scattering potential of the impurity is weak or extended in real space, the intra-pocket scattering may dominate and cannot significantly break the Cooper pairs[44]. Considering that the scattering potential of these defects may be weak, and the measuring temperature is relatively high (0.3K~0.4$T_C$), we cannot rule out the s± paring symmetry from the absence of in-gap states in our measurements. A similar argument should be made for the flat vortex core states here. The vanishing gap at the vortex core indicates a large population of bound states and to distinguish these from Majorana zero mode would require future experiments with much better energy resolution.

**Conclusion**

In summary, we show spectroscopic evidence that the enhanced superconductivity coexists with the characteristic broken-inversion-symmetry $T_d$ phase at the atomic scale, thus allowing the non-trivial band topology. While detailed consequence of heterogeneity is left for future theoretical works, our finding provides experimental support for the Se-doped $MoTe_2$ as a potential candidate for tunable Weyl superconductivity or time-reversal invariant topological superconductor with enhanced superconducting critical temperature Tc.

**Acknowledgements**

Work at the University of Illinois, Urbana-Champaign was supported by U.S. Department of Energy (DOE), Office of Science, Office of Basic Energy Sciences (BES), Materials Sciences and Engineering Division under Award # DE-SC0022101. Z.Y.W. is supported by National Natural Science Foundation of China (No. 12074364) and the Fundamental Research Funds for the Central Universities (WK3510000012). This work was, in part, supported by a JST-CREST project [JPMJCR16F2] and a JSPS Grants-in-Aid for Scientific Research (A) [21H04652].

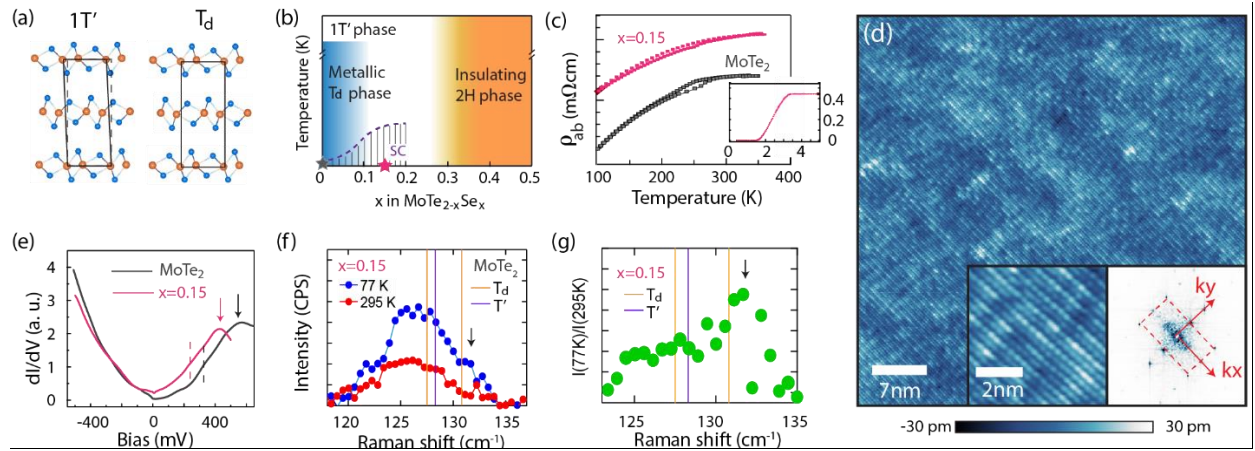

**Figure 1.** (a) Crystal structure of 1T' and $T_d$ phases of $MoTe_2$. The solid dark lines denote one unit-cell. (b) An illustration of phase diagram of $MoTe_{2-x}Se_x$. The dark gray and pink stars denote the two doping level studied in this work (x=0 and 0.15). (c) Resistivity as a function of temperature between 100 K and 350 K for the pristine (dark gray) and Se doped samples (pink), respectively. The anomaly at 250K shows a 1T' to $T_d$ phase transition. The onset of superconducting transition in $MoTe_{1.85}Se_{0.15}$ occurs around 3K as shown in the inset. (d) Typical topographic images taken on the (001) cleaved surface at 0.3 K ($V_S$=100 mV, $I_t$=50 pA). The inset shows an atomically resolved image (10 mV, 50 pA) and the FFT of (d), where the dashed red box indicates the 1st Brillouin zone. The topographic corrugation has a typical length scale of ~20 nm (Fig. S5). (e) *dI/dV* spectra taken on the pristine (dark gray) and Se doped (pink) samples ($V_s$=-500 mV, $I_t$=0.5nA). The dashed lines and arrows mark the features in the density of states that are related to the top of two bulk bands[33,34]. (f) Raman shifts of $MoTe_{1.85}Se_{0.15}$ at 77K and 295K. (g) Normalized Raman spectra I(77K)/I(295K). The yellow and purple lines denote the Raman shifts for $T_d$ and T' phase of $MoTe_2$, respectively.

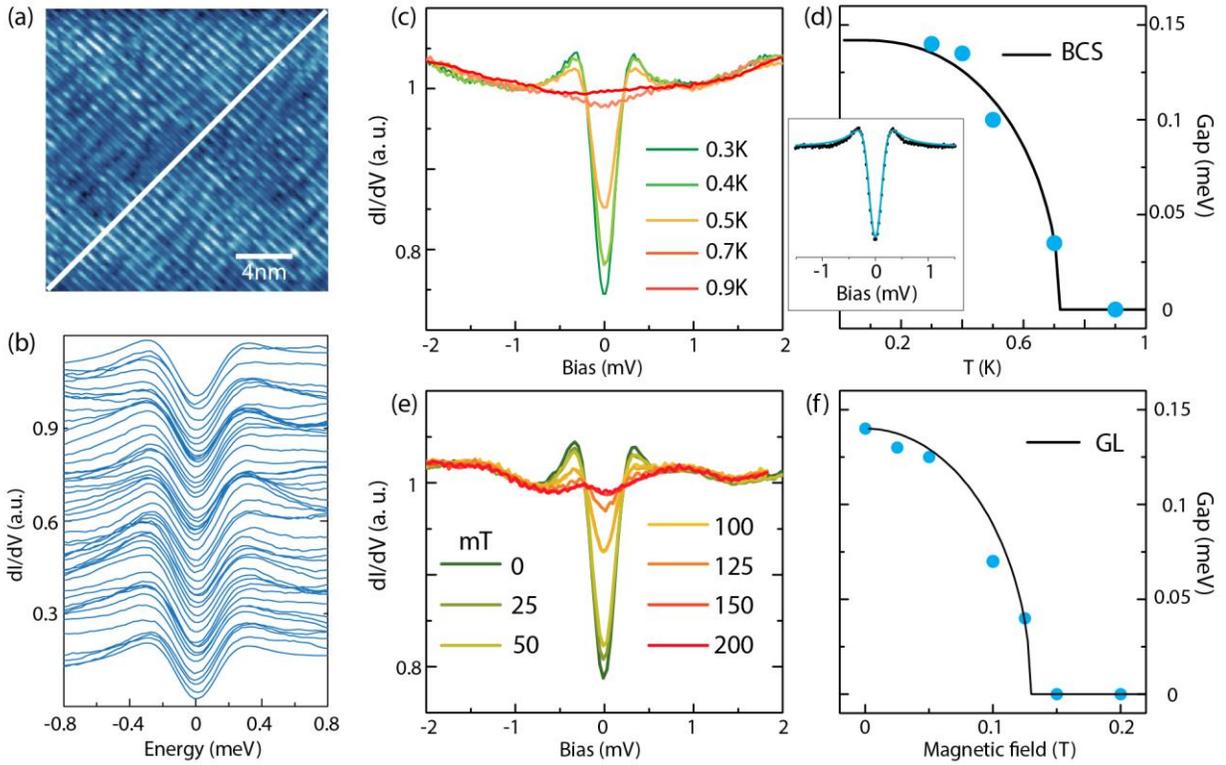

**Figure 2.** (a) Topographic image ($V_S$=10mV, $I_t$=50pA). The white line indicates the trace on which the spectra shown in (b) are obtained. (b) dI/dV spectra taken at 0.3K showing the homogenous superconductivity in the doped sample. (c, d) Temperature dependence of the tunneling spectra and superconducting gap. The inset of (d) shows a BCS fitting of the spectra at 0.3K. (e, f) Magnetic field dependence of the tunneling spectra and the gap. The dark line indicates the expectation of Ginzburg-Landau theory.

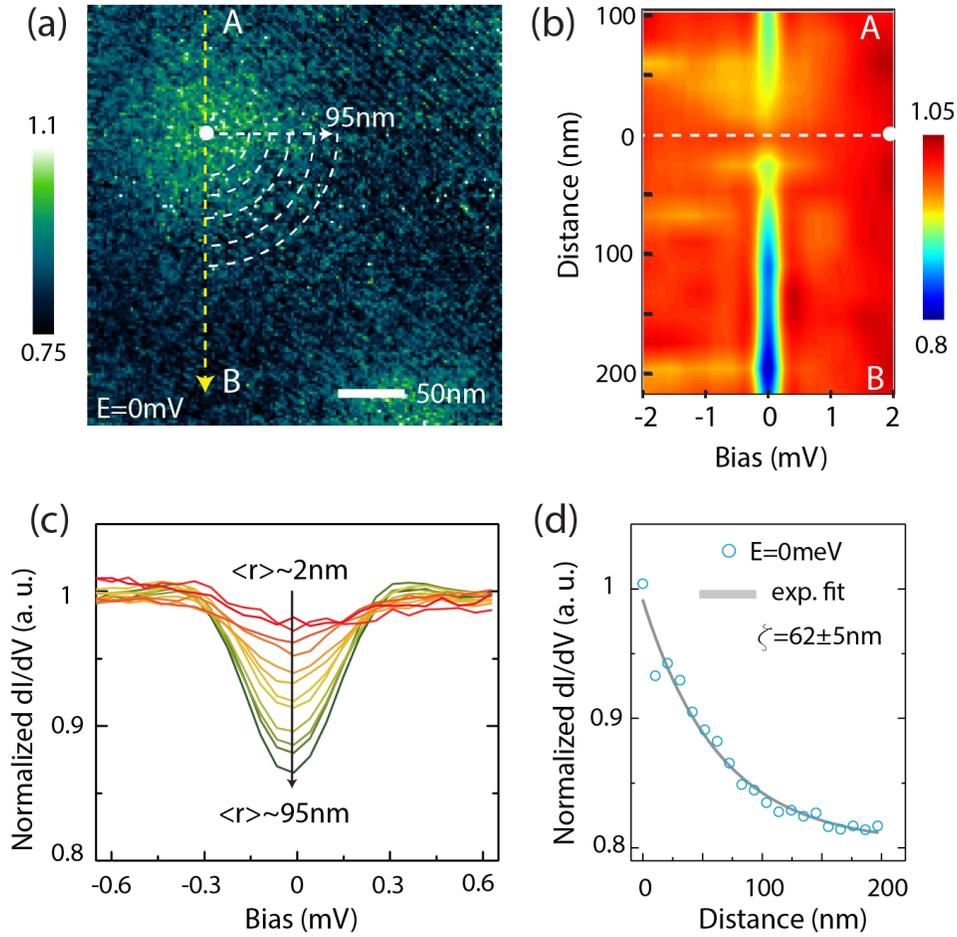

**Figure 3.** (a) Differential conductance map measured with a magnetic field of 0.04T perpendicular to the surface. (b) Spatially resolved tunneling spectra measured across a single vortex along the dashed yellow line (A-B) shown in (a). (c) Tunneling spectra as a function of distance (up to 95nm) from the vortex core, azimuthally averaged for a range of 90 degrees in the bottom-right corner (marked by the white arcs). (d) Distance dependence of the zero-bias conductance (ZBC) from vortex core up to 200 nm and the fit with an exponential decay $g(r) = g_0 + A exp(-r/\xi)$. The data are obtained along the dashed yellow line shown in (a), form the center of the vortex to B point. Data are taken at 0.3K.

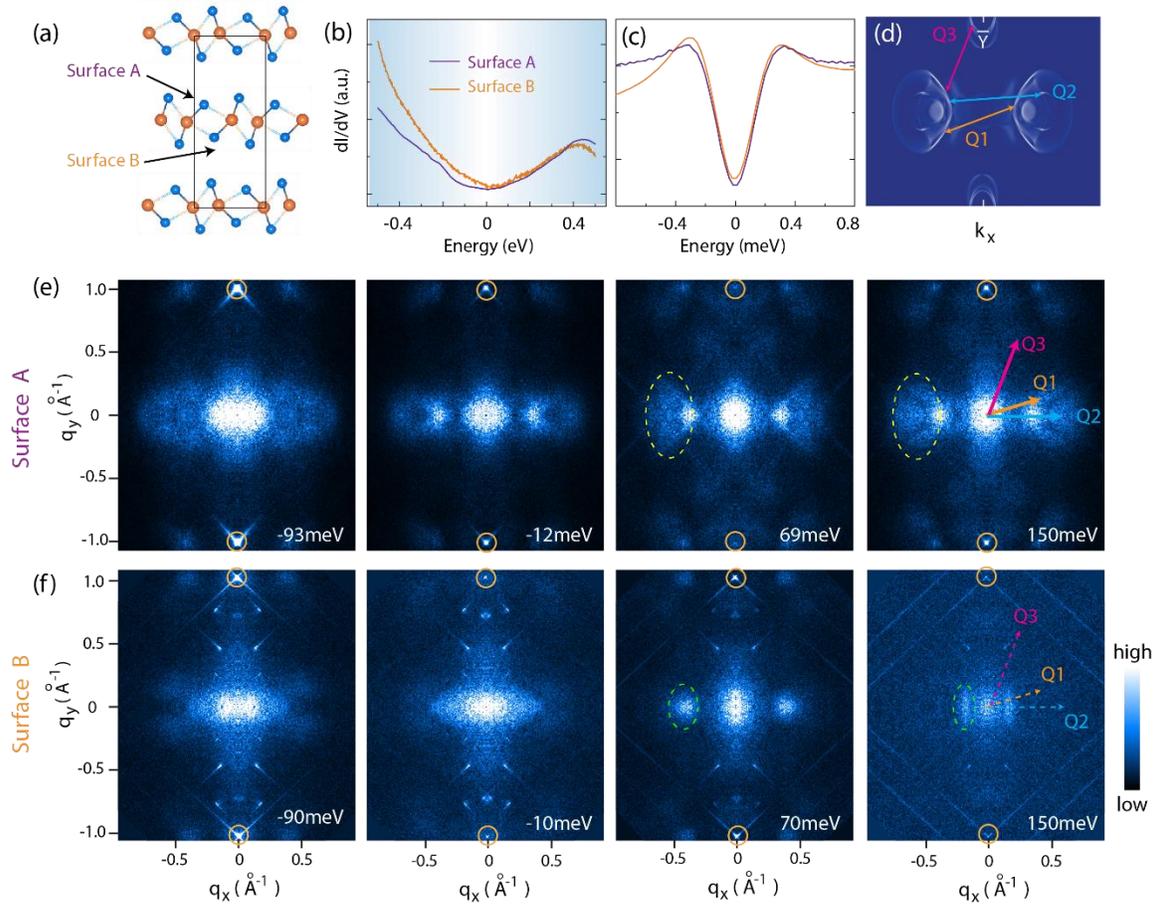

**Figure 4.** Spectroscopic evidence for the broken inversion symmetry in MoTe$_{1.85}$Se$_{0.15}$. (a) The schematic for two inequivalent surfaces obtained by cleaving opposite side of a bulk crystal. (b-c) Tunneling spectra obtained on two different surfaces. (d) Cartoon illustration of the constant energy contours and the related scattering vectors. (e) QPI patterns obtained on surface A at energies of -93, -12, 69, and 150mV, respectively. (f) QPI patterns obtained on surface B at energies of -90, -10, 70 and 150mV, respectively. The images are mirror-symmetrized with respect to K$_x$=0. We have identified multiple scattering channels as follows: Q$_1$, intra-band scattering of the trivial surface states; Q$_2$, scattering between surface states and the electron pockets the bulk electron pockets; Q$_3$, scattering between the surface states and the projected bulk states at the $\bar{Y}$ points. The yellow circles mark the Bragg peaks along K$_y$ direction, and the dashed eclipses highlight the differences in the QPI patterns. Data are taken at 0.3K.

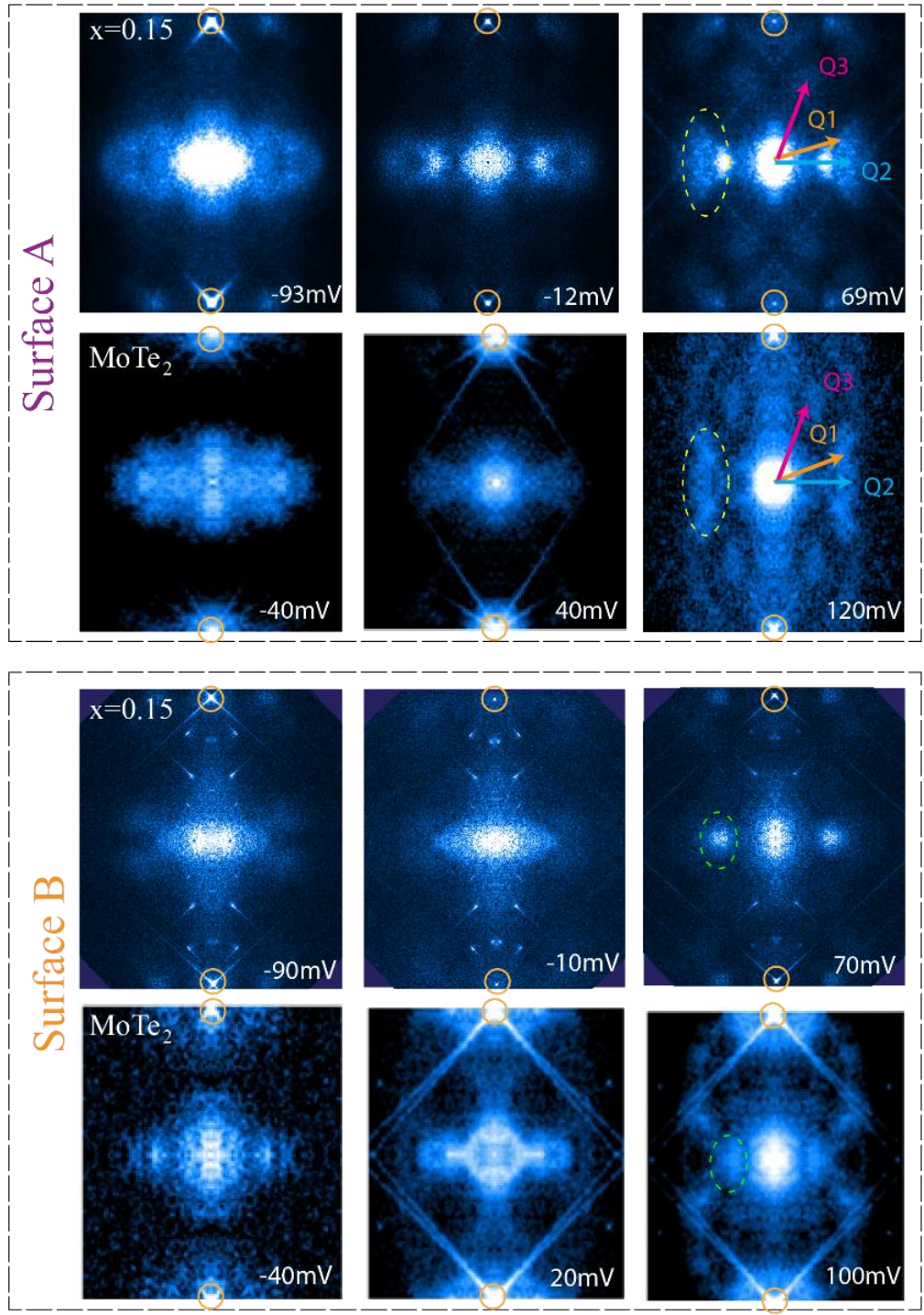

**Figure 5. Comparison between QPI patterns of MoTe$_2$ and MoTe$_{1.85}$Se$_{0.15}$, for both surface A and B**. One can find reasonably well analogy for QPI patterns of both surfaces A and B, by simply shifting the chemical potential between doped and none-dopes samples. The yellow circles mark the Bragg peaks along K$_y$ direction and dashed eclipses highlight the differences in the QPI patterns. Here, the data of pristine MoTe$_2$ is from our previous work [ref. 34].